\documentclass{webofc}
\usepackage[varg]{txfonts} % Web of Conferences font
\usepackage{xcolor}
\newcommand{\hMpc}{{\ifmmode{,h^{-1}{\rm Mpc}}\else{$h^{-1}$Mpc}\fi}}
\newcommand{\hkpc}{{\ifmmode{,h^{-1}{\rm kpc}}\else{$h^{-1}$kpc}\fi}}
\newcommand{\hMsun}{{\ifmmode{\,h^{-1}{\rm {M_{\odot}}}}\else{$h^{-1}{\rm{M_{\odot}}}$}\fi}}
\newcommand{\Msun}{\,\rm {M_{\odot}}}
\newcommand{\Mstar}{{\ifmmode{,M_{*}}\else{$M_{*}$}\fi}}
\newcommand{\Mhalo}{{\ifmmode{\,M_{\rm halo}}\else{$M_{\rm halo}$}\fi}}
\newcommand{\ltsima}{$\; \buildrel < \over \sim \;$}
\newcommand{\gtsima}{$\; \buildrel > \over \sim \;$}
\newcommand{\lsim}{\lower.5ex\hbox{\ltsima}}
\newcommand{\gsim}{\lower.5ex\hbox{\gtsima}}

\newcommand{\theth}{{\sc The Three Hundred}}

\newcommand{\gadgetx}{\textsc{Gadget-X}}

\newcommand{\MCNN}{M_{\text{pred}}}
\newcommand{\Mtrue}{M_{\text{true}}}
\newcommand{\MPlanck}{M_{\text{Planck}}}

\newcommand{\thethreehundred}{{\sc The Three Hundred}}
\begin{document}
\title{Mass Estimation of Planck Galaxy Clusters using Deep Learning}

\author{\firstname{Daniel} \lastname{de Andres}\inst{1}\fnsep\thanks{\email{daniel.deandres@uam.es}} \and 
	\firstname{Weiguang} \lastname{Cui}\inst{2}
	\and
	\firstname{Florian} \lastname{Ruppin}\inst{3}
	\and 
	\firstname{Marco}
	\lastname{De Petris}\inst{4}
	\and
	\firstname{Gustavo}
	\lastname{Yepes}\inst{1}
	\and 
	\firstname{Ichraf}
	\lastname{Lahouli}\inst{5}
	\and
	\firstname{Gianmarco}
	\lastname{Aversano}\inst{5}
	\and
	\firstname{Romain}
	\lastname{Dupuis}\inst{5}
	\and
	\firstname{Mahmoud}
	\lastname{Jarraya}\inst{5}
        % etc.
}

\institute{ Departamento de Física Teórica  M-8 and  CIAFF, Universidad Autónoma de Madrid, Cantoblanco 28049, Madrid, Spain
\and
          Institute for Astronomy, University of Edinburgh, Blackford Hill, Edinburgh, EH9 3HJ, UK
\and    
         Kavli Institute for Astrophysics and Space Research, Massachusetts Institute of Technology, Cambridge, MA 02139, USA
\and
          Dipartimento di Fisica, Sapienza Universit\'a di Roma, Piazzale Aldo Moro, 5-00185 Roma, Italy
\and
        EURANOVA, Mont-Saint-Guibert, Belgium
        }

\abstract{%
    Galaxy cluster masses can be inferred indirectly using measurements from X-ray band, Sunyaev-Zeldovich (SZ) effect signal or optical observations. Unfortunately, all of them are affected by some bias. Alternatively, we provide an independent estimation of the cluster masses from the Planck PSZ2 catalog of galaxy clusters using a machine-learning method. We train a Convolutional Neural Network (CNN) model with the mock SZ observations from \thethreehundred\ (the300) hydrodynamic simulations to infer the cluster masses from the real maps of the Planck clusters. The advantage of the CNN is that no assumption on a priory symmetry in the cluster's gas distribution or no additional hypothesis about the cluster physical state are made. We compare the cluster masses from the CNN model with those derived by Planck and conclude that the presence of a mass bias is compatible with the simulation results.
}
\maketitle
\section{Introduction}
\label{intro}
Galaxy clusters are the biggest gravitational bound objects in the Universe and  their number density as a function of their mass and redshift is very sensitive to the matter content of the Universe and the expansion history. Therefore, they constitute one of the best cosmological probes to constrain different cosmological parameters  \cite{planck2020}. However, the total mass of a cluster is not a direct measurable quantity. It is usually inferred from several observables under some physical assumptions. An accurate determination of the total mass of galaxy clusters is in fact a key problem in physical cosmology.

Among the different cluster mass definitions, we are considering $M_{500}$ defined as the total mass contained in a sphere of radius $R_{500}$, within which the density is 500 times the critical density:
\begin{equation}
    M_{500} = \frac{4\pi}{3}R^{3}_{500}500\rho_{\text{c}}(z),
\end{equation}
here $z$ is the redshift and $\rho_{\text{c}}(z) \equiv 3H(z)^2 /8\pi G $ is the critical density.

By observing the millimeter wavelength sky one can measure the spectral distortions of the cosmic microwave background (CMB) photon due to the inverse-Compton scattering by the free electrons of the intra cluster medium (ICM), i.e. the Sunyaev-Zeldovich (SZ) effect \cite{SZeffect}. Through the SZ effect, thousands of clusters have been observed, and discovered, by the South Pole Telescope (SPT; \cite{SPT}), the Atacama cosmology telescope (ATC; \cite{ATC}), and the Planck telescope \cite{planckszcatalog}. The intensity of the thermal component of the SZ effect (tSZ) over the line of sight is quantified by the Compton parameter $y$, defined as
\begin{equation}\label{eq:defy}
    y= \frac{\sigma_{\text{T}}k_{\text{B}}}{m_{\text{e}}c^{2}}\int n_{\text{e}}T_{\text{e}}dl \text{ ,}
\end{equation}
where $\sigma_{\text{T}}$ is the Thomson cross section, $k_{\text{B}}$ is the Boltzmann constant, c the speed of light, $m_{\text{e}}$ the electron rest-mass, $n_{\text{e}}$ the electron number density, $T_{\text{e}}$ the gas temperature.

One procedure to derive the total mass of the cluster from the $y$-Compton parameter map is using a scaling relation $Y_{500}-M_{500}$, where $Y_{500}$ denotes the integrated value of the $y$ parameter within $R_{500}$. 

Nevertheless, in the Planck catalog, the calibration of this scaling relation is carried out assuming hydrostatic equilibrium. Therefore the estimated mass , $M_{\text{est}}$, is expected to be biased. This bias parameter $b$, is usually defined as $1-b = M_{\text{est}}/M_{\text{true}},$
where $M_{\text{true}}$ is the total mass of the clusters. The determination of cosmological parameters, such as the matter density $\Omega_{\text{m}}$ and the present amplitude of density fluctuations $\sigma_{8}$ is sensitive to the particular value of $b$. Particularly, the bias value $b$ needed to find no tension with CMB observations is $(1-b)=0.58\pm 0.04$ \cite{planckbiascmb}, or the latest result $(1-b)=0.62\pm 0.05$ \cite{salvaticmb}. However, the bias value is about $1-b\simeq0.8$ from numerical simulations, which is clearly in a disagreement with CMB observations (e.g. \cite{makiyabias, giulia}). In addition, recent analyses of weak lensing observations led to a value of the bias $1-b=0.84\pm 0.04$ \cite{herbonetwl}, which is in agreement with simulations.

In recent development, machine learning techniques have been used in simulations to provide an unbiased estimation of the cluster total mass by training the algorithms using mock observation in X-ray frequencies \cite{ntampaka}, simulated SZ observations \cite{guptasz}, and in a multi-channel approach combining X-ray, SZ and stellar mock maps \cite{bahamas}.

In this work, we aim at predicting the total mass of real galaxy clusters selected from the Planck Compton-$y$ parameter maps by training a Convolutional Neural Network (CNN) on simulated maps extracted from a large dataset of hydrodynamic simulations provided by \theth\ collaboration \citep{threehundred}.

\section{Data set}
\label{sec-1}

The mock Compton-$y$ parameter maps were computed from the results of the \theth\ hydrodynamic simulations. They consist on the resimulations of spherical regions of radius $15 h^{-1} Mpc$ centered around the most massive 324 clusters found at $z=0$ MULTIDARK dark-matter-only simulation (MDPL2; \cite{MDPL2}) with cosmological parameters from the Planck collaboration \cite{planckparameters}. Particularly, in this work we have used the results from the $\gadgetx$ \cite{murantegadgetx,rasiagadgetx} runs. The cluster size halos are identified with the AHF algorithm \cite{AHF}. Then, we have chosen halos with masses greater than $10^{14}\hMsun$ at redshift $z<1$. Moreover, the tSZ signal corresponding to these clusters is therefore computed using equation \ref{eq:defy} with the {\sc PYMSZ} public software \cite{PYMSZ}. The images are generated to have a resolution of $1920\times 1920$ pixels with a fixed angular resolution of 5'' so that all images will cover at least $R_{200}$ for all clusters. 

A Gaussian smoothing is then applied to these maps in order to mimic the impact of the Planck Beam (FWHM = 10 arcmin) and we further add a similar statistical noise to mimic the real Planck observations. In summary, the simulated data set consists on 7016 selected clusters with 27 random rotations each, amounting to a total of  $191,862$ maps. They have been particularly selected in order to cover the same mass and redshift range of the selected Planck cluster sample. The real Planck cluster catalog consists of $1,096$ observed tSZ maps selected from the Second Planck Catalogue of Sunyaev-Zeldovich Sources (PSZ2; \cite{PLSZ2}) with known redshifts.  In figure 1, we show the Probability Distribution Function (PDF) corresponding to the distribution in redshift and mass of the simulated Planck clusters and real Planck clusters used in this work. An example of the $y$-maps for the simulated clusters is presented in figure 2.

\begin{figure}[h]
\begin{center}  
\includegraphics[scale=0.5]{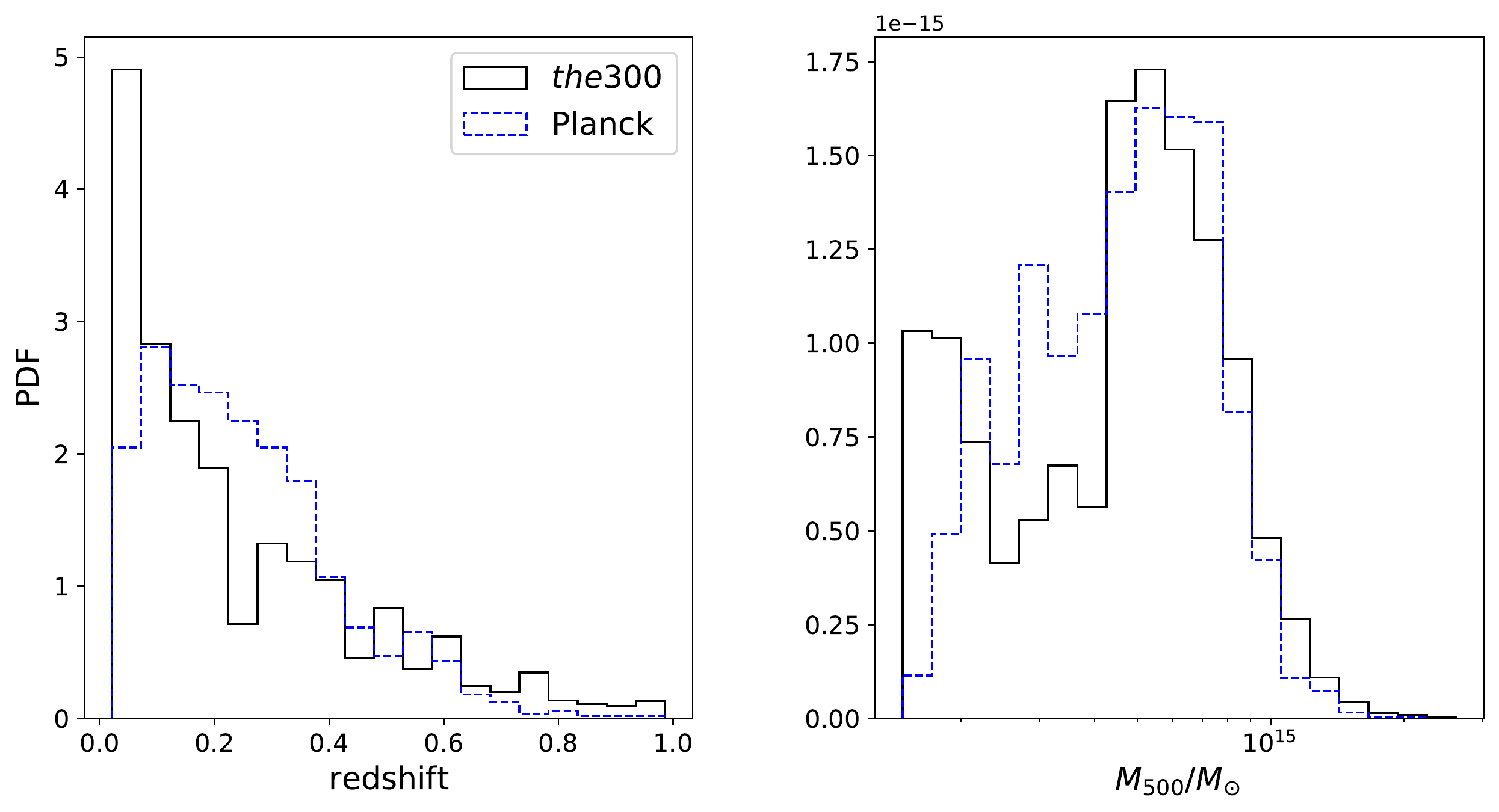}
\caption{PDF of the cluster samples ( \thethreehundred{} simulation in solid black line and Planck PSZ2 clusters in blue dashed line) corresponding to the distribution in redshift (left panel) and mass (right panel).}\label{fig-1}
\end{center}
\end{figure}

\begin{figure}[h]
\begin{center}  
\includegraphics[scale=0.55]{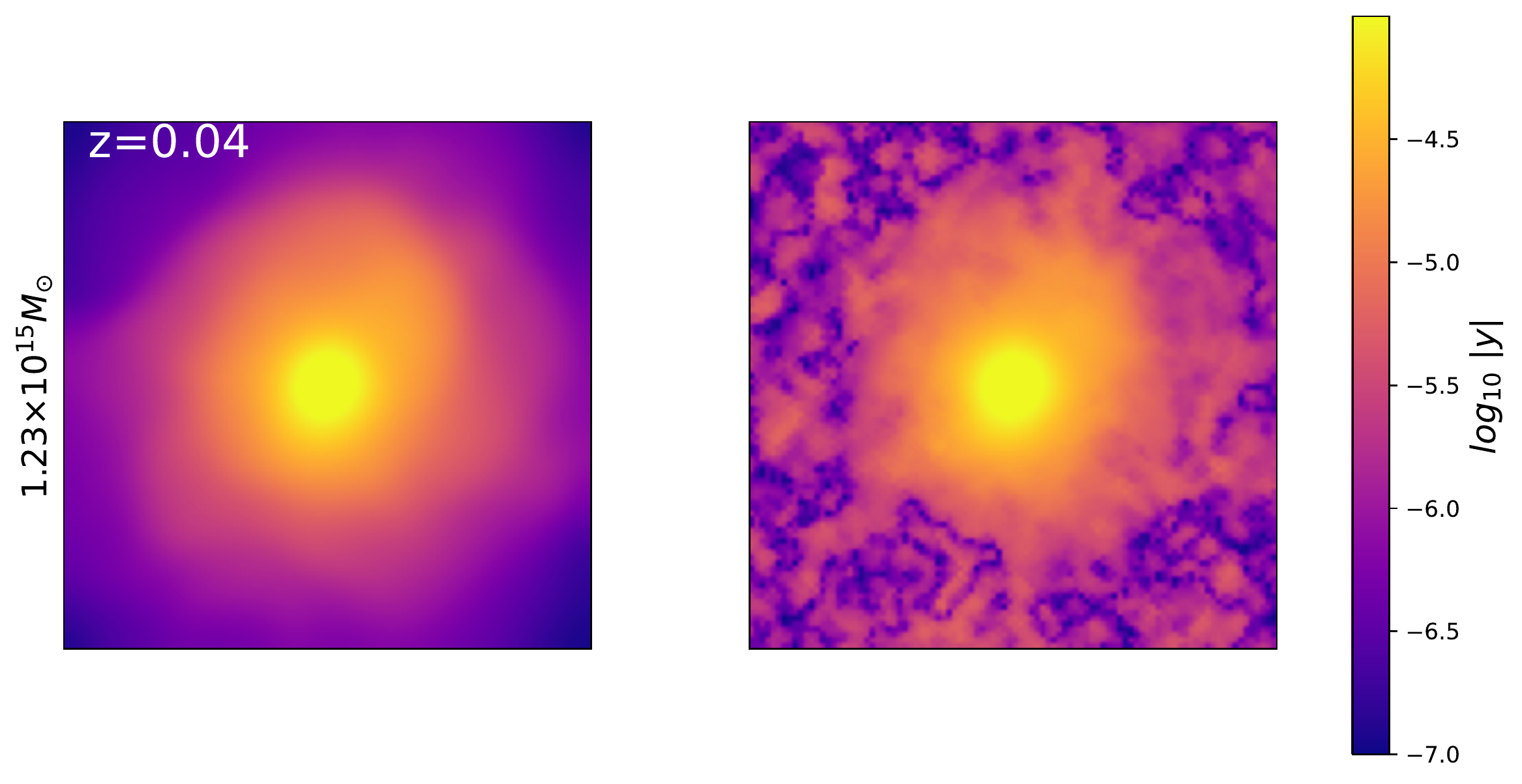}
\caption{Example of simulated tSZ maps. The left panel corresponds to the clean map and the right panel represents the same image with Planck instrumental beam size and noise. The value of this cluster mass is $M_{500}=1.23\times10^{15}\Msun$ and the corresponding redshift is $z$=0.04. In addition, the color bar shows the logarithmic value of the Compton-$y$ parameter.}\label{fig-2}
\end{center}

\end{figure}

\section{Algorithm}
In this work, we train a deep learning model (e.g. \cite{deeplearningbook}) using the mock images of our simulated cluster sample. Our particular model is based on a CNN model, that can find a mapping from a two dimensional image $I_{ij}$ to a scalar quantity, i.e. $M_{500}$. Our CNN architecture is a simplified version of the architecture proposed by Simonyan \& Zizzerman commonly known as {\sc VGGNet} \cite{vggnet}. This architecture has already been used to infer cluster masses from simulated observations in  \cite{ntampaka} and \cite{bahamas}. %Therefore, we utilize a similar CNN model.
Furthermore, we have checked that more dense and complicated VGGNet architectures reproduce similar results based on our data set. To train our algorithm, we consider the following loss function
%logarithmic mean squared error is used error is considered as our loss function, i.e.
%
\begin{equation}
    \mathcal{L}=\frac{1}{N}\sum_{i=1}^{N}(\log \Mtrue^{i} -\log \MCNN^{i})^{2}\text{ ,}
\end{equation}
where $\MCNN^{i}$ is the predicted CNN mass for cluster $i$ and $N$ is the total number of cluster images in the training set.  We then split our data set in 80\% training, 10\% validation, and 10\% test. The validation data set is used to ensure that the model produces an unbiased estimation of the mass. The test set is used exclusively to asses the final accuracy of the model. We have observed that mixing all the redshifts together yields to a very poor performance. Therefore, we have trained 4 CNNs by diving the data set in 4 different redshift bins: $z\leq0.1$ ; $0.1< z\leq 0.2$ ; $0.2<z\leq 0.4$ and $0.4>z$.

% In addition, we have observed that mixing all the redshifts together yields to a very poor performance.
% Therefore, we have trained four CNNs by dividing the data set in 4 different redshifts bins: $z\leq0.1$ ; $0.1< z\leq 0.2$ ; $0.2<z\leq 0.4$ and $0.4>z$.

Only open source python libraries have been used in this project. Particularly, {\sc Keras} \cite{keras} with {\sc Tensorflow} \cite{tensorflow} GPU acceleration.

\section{Results}
\label{sec-3}

\begin{figure}[h]
\begin{center}  
\sidecaption
\includegraphics[width=8cm,clip]{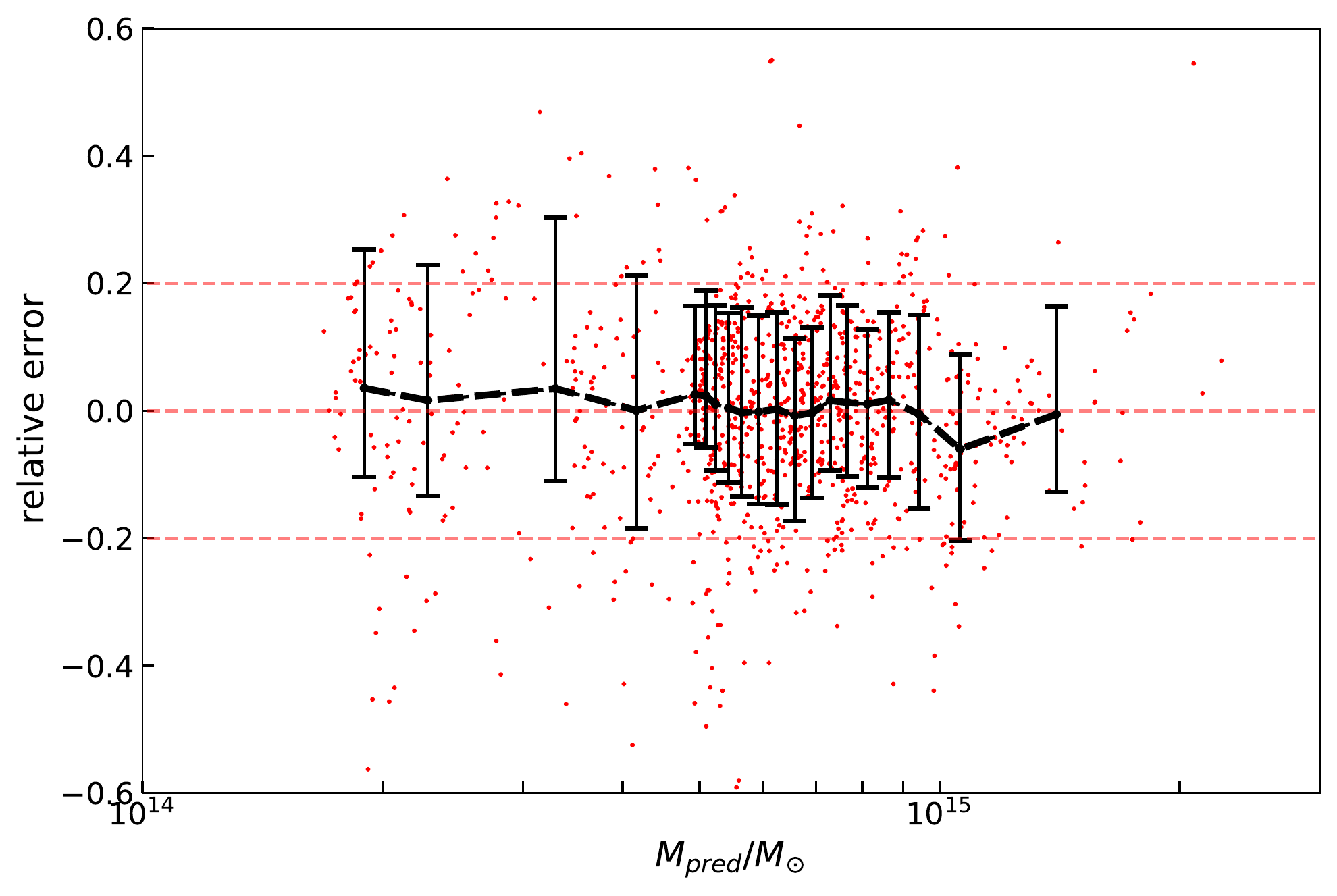}
\caption{The value of the relative error $(\MCNN-\Mtrue)/\MCNN$ as a function of the predicted mass $\MCNN$ for \theth\ mock data. The dashed black line represents the median for each mass bin while the error bars represents the 68\% confident interval. Furthermore, red dashed lines represent the 0 relative error line and $\pm20\%$ error. Only 1000 points (in red) drawn randomly from the original distribution are displayed. The error bars are computed using the full test set ($\simeq 20000$ maps).}\label{fig-3}
\end{center}

\end{figure}

\begin{figure}[h]
\begin{center}  
\sidecaption
\includegraphics[width=8cm,clip]{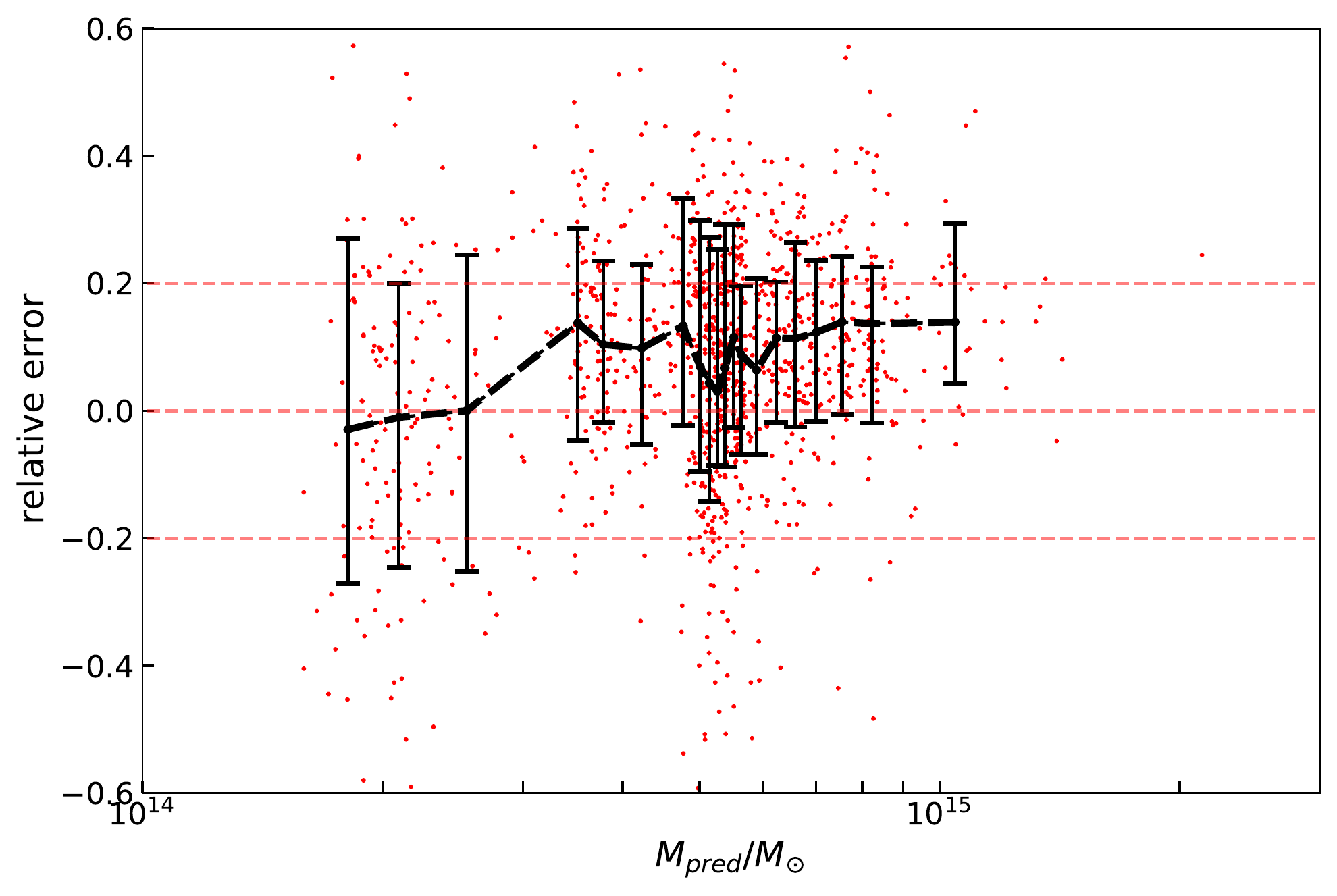}
\caption{The value of the relative error $(\MCNN-\MPlanck)/\MCNN$ as a function of the predicted mass $\MCNN$ for the \textbf{real Planck data set}. The error bars and lines are the same as figure \ref{fig-3}.}\label{fig-4}
\end{center}
\end{figure}

After training the CNN models with Planck-like simulated Compton-$y$ parameter maps, we predict the masses of the PSZ2 cluster catalog. In order to study the performance of these predictions, we compute the relative error defined as the relative difference of the Planck $\MPlanck$ mass and the predicted CNN mass $\MCNN$: $\text{relative error}=\frac{M_{CNN}-M_{Planck}}{M_{CNN}}\text{.}$

In figure \ref{fig-3}, we show the relative error $(\MCNN-\Mtrue)/\MCNN$ as a function of the predicted mass $\MCNN$ using simulated data. According to our results depicted in the figure \ref{fig-3}, the relative error with respect to the 3D dynamical simulated mass is always smaller than 5\% for this considered mass range and the scatter (standard deviation $\sigma$) is 17\%. A similar result by applying the same CNN to real Planck maps is shown in figure~\ref{fig-4}. However, note that the CNN mass is an unbiased estimator of the cluster mass as shown in figure~\ref{fig-3}, while the Planck cluster mass show systematic difference to the CNN mass. The bias parameter $b$ can then be computed as the mean of relative difference between $\MCNN$ and $\MPlanck$: $b\simeq(\MCNN-\MPlanck)/\MCNN$. We would like to highlight that the relative error is biased $1-b=0.86$ for massive clusters $\MCNN>4\times 10^{14}\Msun$. However, in the low mass range the CNN predictions seem unbiased with respect to the masses estimated by Planck. 

A possible explanation for that is by taking into consideration that Planck collaboration  \cite{planck2014} calibrated the scaling relation from a mass-proxy relation \cite{kravtsov} with a slope of 1.79, whereas the $\gadgetx$ clusters of \thethreehundred{} simulations is in agreement with a self-similar scaling relation \cite{threehundred} with a slope of $5/3$.

\section{Conclusion}
\label{sec-4}

In this work, we have estimated the mass of clusters by exploring Compton-$y$ parameter maps provided by Planck. In order to do so, we have created a catalog of almost $200,000$ mock Compton-$y$ maps mimicking the noise levels of Planck observations. Then, a CNN is trained with the mock images aiming at predicting the mass of real Planck observations.

Our results show the presence of a mass bias $b$ dependent on cluster mass. The masses estimated by Planck are unbiased with respect to our predicted masses for low mass clusters, while the bias for massive clusters is $1-b=0.86^{+0.12}_{-0.7}$. 

Otherwise, in hydrodynamic simulations, the value of this bias does not depend on the cluster mass \cite{giulia} and its value is of order 10-20\%. This is in agreement with the fact that the parameters of \theth\ $Y_{500}-M_{500}$ scaling relation are compatible with the self-similar relation. 

\section*{Acknowledgments}
We would like to thank the Red Espa\~nola de Supercomputaci\'on for granting access to the Marenostrum supercomputer where most of \theth\ simulations have been performed.
GY and DA would like to thank  MICIN/Feder (Spain) for partial financial support under project grant PGC2018-094975-C21. WC is supported by the STFC AGP Grant ST/V000594/1 and the science research grants from the China Manned Space Project with NO. CMS-CSST-2021-A01 and CMS-CSST-2021-B01. MDP  acknowledges support from Sapienza Università di Roma thanks to Progetti di Ricerca Medi 2020, prot. RM120172B32D5BE2.

% For tables use syntax in table~\ref{tab-1}.
% \begin{table}
% \centering
% \caption{Please write your table caption here}
% \label{tab-1}       % Give a unique label
% % For LaTeX tables you can use
% \begin{tabular}{lll}
% \hline
% first & second & third  \\\hline
% number & number & number \\
% number & number & number \\\hline
% \end{tabular}
% % Or use
% \vspace*{5cm}  % with the correct table height
% \end{table}
%
% BibTeX or Biber users please use (the style is already called in the class, ensure that the "woc.bst" style is in your local directory)
% \bibliography{name or your bibliography database}

\end{document}